\documentclass{mn2e}

\usepackage{epsfig}

\title{The quiescent light curve and evolutionary state of GRO J1655--40}

\author[M. E. Beer and Ph.\ Podsiadlowski]{Martin
E. Beer\thanks{E-mail: beer@astro.ox.ac.uk} and Philipp Podsiadlowski
\\University of Oxford, Nuclear and Astrophysics Laboratory, Oxford,
OX1 3RH, England}

\date{\today}

\volume{000}

\pagerange{\pageref{firstpage}--\pageref{lastpage}} \pubyear{2001}

\begin{document}

\maketitle

\label{firstpage}

\begin{abstract}
We present ellipsoidal light-curve fits to the quiescent $B$, $V$, $R$
and $I\/$ light curves of GRO J1655--40 (Nova Scorpii 1994). The fits
are based on a simple model consisting of a Roche-lobe filling
secondary and an accretion disc around the black-hole primary. Unlike
previous studies, no assumptions are made about the interstellar
extinction or the distance to the source; instead these are determined
self-consistently from the observed light curves. In order to
obtain tighter limits on the model parameters, we used the distance
determination from the kinematics of the radio jet as an additional
constraint.  We obtain a value for the extinction that is lower than
was assumed previously; this leads to lower masses for both the black
hole and the secondary star of 5.4 $\pm$ 0.3 M$_{\sun}$ and 1.45 $\pm$
0.35 M$_{\sun}$, respectively. The errors in the determination of the
model parameters are dominated by systematic errors, in particular due
to uncertainties in the modeling of the disk structure and
uncertainties in the atmosphere model for the chemically anomalous
secondary in the system.  A lower mass of the secondary naturally explains the
transient nature of the system if it is either in a late case A or
early case B mass-transfer phase.

\end{abstract}


\section{Introduction}




GRO J1655--40 is a well studied soft X-ray transient (SXT) with an
orbital period of $2.62168 \,\pm\, 0.00014\,$d (van der Hooft et al.\
1997, hereafter vdH). It contains a F3--5 giant or sub-giant with an
effective temperature of approximately 6500\,K (Orosz \& Bailyn 1997,
hereafter OB). Shahbaz et al.\ 1999 (hereafter S99) obtained the
radial-velocity curve of the system during quiescence using
high-resolution spectroscopy and found a mass function of $2.73 \,\pm\,
0.09\,$M$_{\sun}$. This value is significantly lower than the previous
estimates by Bailyn et al.\ (1995b, $3.16 \,\pm\, 0.15\,$M$_{\sun}$) and
OB ($3.24 \,\pm\, 0.09\,$M$_{\sun}$), most likely because their
radial-velocity determination relied on some data that were obtained
during outburst (see Phillips, Shahbaz \& Podsiadlowski 1999). 
From the rotational broadening of the spectral lines, they also
obtained a constraint on the mass ratio of 2.29\,--\,3.06.
 
In their original study, OB modelled the ellipsoidal light curve during
quiescence assuming a polar temperature of 6500~K for the secondary
and found an inclination of $69\fdg 50\,\pm\, 0\fdg 08$ and a mass
ratio of $2.99 \,\pm\, 0.08$.  These values imply masses of $7.02 \,\pm\,
0.22\,$M$_{\sun}$ and $2.34 \,\pm\, 0.12\,$M$_{\sun}$ for the black hole
and secondary star, respectively. VdH also modelled the ellipsoidal
light curve during quiescence and obtained an inclination of
$63\fdg 7 - 70\fdg 7$ and a secondary mass in the range of
1.60\,--\,3.10\,M$_{\sun}$, which implies a mass ratio of
2.43\,--\,3.99.

VdH's estimates have larger uncertainties since, unlike OB, they
considered models with three different luminosities (31, 41 and 
54\,L$_{\sun}$) taking into account uncertainties in the distance ($3.2\,\pm\,
0.2\,$kpc, obtained from the kinematics of the observed radio jet;
Hjellming \& Rupen 1995) and in the colour excess ($E(B-V)= 1.3\,\pm\,
0.1$, based on various previous estimates; see vdH). Combining a
value for $E(B-V)$ of 1.3 with the observed $B-V$ colour of approximately
1.55 implies an intrinsic $(B-V)_0$ of less than 0.25 (this is an
upper limit, since any disc contribution tends to be redder than
this). A $(B-V)_0$ of less than 0.25 corresponds to a sub-giant of
spectral type A8 or earlier (Fitzgerald 1970) whilst the $(B-V)_0$ of
a F3--5 giant or sub-giant is 0.39\,--\,0.44. This immediately
demonstrates that a value for $E(B-V)$ of 1.3 is not consistent with
the observed spectral type. An overestimate of $E(B-V)$ leads to an
overestimate of the bolometric luminosity of the secondary, which in
turn requires a larger and more massive secondary. More recent
estimates of the ultraviolet extinction (Hynes et al.\ 1998) yielded 
an $E(B-V)$ of $1.2 \,\pm\, 0.1$. This implies a secondary of lower
luminosity than considered by vdH.

Both OB and vdH allowed an arbitrary magnitude offset for each
passband ($B$, $V$, $R$, $I$) when they modelled the ellipsoidal
light curves. This has the same effect as allowing the distance and the
colour excess to vary independently for different passbands. It also means
that they were not using all the available information. In particular,
this did not allow them to check whether their best-fitting models were 
actually consistent with the observed spectral type.

In the present study we avoid these problems by fitting the
ellipsoidal light curves for all passbands simultaneously (without
arbitrary offsets), but allowing the distance and the colour excess
to vary freely. As a consequence, the distance and the colour excess are 
determined self-consistently from the best-fitting models instead
of being taken as input parameters.

In Section~\ref{modeldescript} we outline the basic model used in this
study.  In Section~\ref{lcmodel} we apply it to fit the quiescent
light-curve data of OB for different disc structures and obtain a new
model for all parameters of GRO J1655--40 and examine their
uncertainties.  In Section~\ref{sectemp} we show how the variation in
temperature across the surface of the secondary limits the accuracy to
which the spectral type of the secondary can be determined. Finally,
in Section~\ref{discuss} we compare our results to previous studies
and discuss the implications of our new model for the evolutionary
state of the system.

\section{Description of the model} \label{modeldescript}

Our method to model the ellipsoidal light curve of Nova Sco is
similar in many respects to the methods used previously by OB and
vdH. It consists of a Roche-lobe filling secondary and a simple
model for the accretion disc. 

For a given value of the polar temperature, the temperature across
the secondary is determined using a standard linear limb-darkening
law, where the coefficients for each surface element are calculated
by interpolating the tables of Wade \& Rucinski (1985).

Similar to OB and vdH, we model the accretion disc as a flat cylindrical 
disc with an opening angle of 2\degr. For the inner
disc radius we follow OB and take, at least initially, a small value 
of 0.005 of the effective Roche-lobe radius ($r_{\rm{L}}$).
We varied the outer disc radius, which is limited by
the tidal disruption radius, considering outer disc radii of 0.7, 0.8
and 0.9 $r_{\rm{L}}$, respectively. For the temperature profile we
adopted a simple power-law profile, $T_{\rm disc}\propto r^{\alpha}$,
where we considered both a flat temperature profile with $\alpha =
-0.1$ and a standard steady-disc profile with $\alpha=-0.75$ (Shakura
\& Sunyaev 1973). The constant of proportionality in the temperature
power-law profile depends on both the outer disc temperature and the
outer disc radius. In order to make the temperature structure
independent of $r_{\rm{out}}$, we scaled the disc rim temperature to
correspond to the same temperature at $r = 0.9\,r_{\rm{L}}$
(i.e. discs with different outer radii will all have 
the same temperature at $r$ = 0.7 $r_{\rm{L}}$).

To model the emission from the secondary and the disc, we divide the
secondary and the surface of the accretion disc into discrete elements
and determine the local gravity and temperature for each
element. Unlike previous studies, we do not assume that the emission on
the secondary is that of a blackbody; instead we calculate it from
the model atmospheres by Kurucz (1992).  For this purpose, we have
constructed a table, which gives the emission in different passbands
($B$, $V$, $R$, $I$) as a function of temperature, surface
gravity and $E(B-V)$.  To calculate the emission, we first corrected the
Kurucz model atmospheres for interstellar extinction for a specified
value of $E(B-V)$ using a mean Galactic extinction curve (Fitzpatrick
1999) and calculated the extinction at each wavelength. It is
important to correct for the extinction first since there is a large
variation in extinction across the broadband passbands: for example in
the $V$ band, $E(\lambda-V)/E(B-V)$ varies from 3.8 to 2.1. We then
folded the extinction-corrected atmospheres through standard filter
response curves (Bessell 1990) to obtain the emission in each
passband. Standard Vega fluxes (T\" ug, White \& Lockwood
1977) were used to calculate the zero point corrections to the
magnitudes. The emission from the disc was calculated similarly
except that extinction-corrected blackbody spectra were folded
through the passbands rather than model atmospheres. Recently Greene,
Bailyn \& Orosz (2001) have also modelled the ellipsoidal light curves
using model atmosphere fluxes and their analysis is discussed further
in Section~\ref{greenecomp}.

The model atmospheres we used in this analysis were of solar
abundance. The metal abundances of the secondary in GRO J1655--40 have
been measured by Israelian et al.\ (1999) who found [Fe/H]\,$=0.1
\,\pm\, 0.2$, but with an overabundance of $\alpha$-elements compared
to solar values. Their $\alpha$-element abundances are [O/H]\,$=1.0
\,\pm\, 0.3$, [S/H]\,$=0.75 \,\pm\, 0.2$, [Mg/H]\,$=0.9 \,\pm\, 0.40$,
[Si/H]\,$=0.9 \,\pm\, 0.3$, [Ti/H]\,$=0.9 \,\pm\, 0.4$ and
[N/H]\,$=0.45 \,\pm\, 0.5$. A solar Fe abundance is supported by
Buxton \& Vennes (2001) who find [Fe/H]\,$=-0.25$\,--\,0.00. As
detailed $\alpha$-element-enhanced model atmospheres are not freely
available we are unable to use $\alpha$-element-enhanced model
atmospheres in our calculations. The majority of the spectral lines,
however, are due to Fe and so the use of the correct Fe abundance is
the most important factor in choosing the metallicity and reassures us
that this is the correct metallicity to choose.

\begin{figure}
\begin{center}
\epsfig{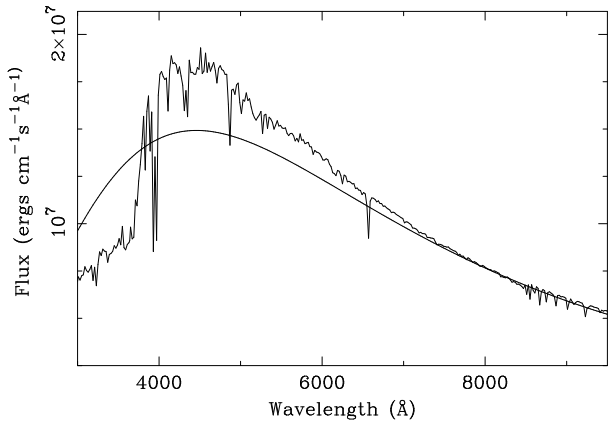}
\caption{The difference in spectra between a 6500 K blackbody and a
model atmosphere of the same temperature and a gravity of 3.0 dex.}
\label{spectra}
\end{center}
\end{figure}

In Fig.~\ref{spectra} we demonstrate how important it is to use model
atmospheres rather than blackbody spectra. It shows the difference
between a 6500 K blackbody and a model atmosphere of the same
temperature with a gravity of 3.0 dex. This temperature corresponds to
a star with a spectral type similar to GRO J1655--40.  The spectra are
significantly different, especially in the $B$ band region.  For the
$I$ band region, the model atmosphere is similar to the Rayleigh-Jeans
tail of a blackbody; hence the blackbody assumption would be valid in
this passband. Orosz \& Hauschildt (2000) have investigated the
difference between blackbodies and model atmosphere calculations and
found that, with the inclusion of model atmospheres, the minima of the
light curve tend to be deeper for the visible and infrared passbands,
since at the minima the coolest parts of the secondary are visible
(e.g. the L1 point at phase 0.5) and for cool temperatures the
differences between model atmospheres and blackbodies are largest. 

For a particular disc model, our ellipsoidal light-curve model has five
free model parameters: the mass ratio, system inclination, colour
excess, distance to Nova Sco and polar temperature ($T_{\rm{pole}}$) of
the secondary star. The individual component masses ($M_1$ and $M_2$)
are calculated from the mass function, where we use the value of $2.73
\,\pm\,0.09\,$M$_{\sun}$ (S99). We determine these five free parameters by
comparing our model ellipsoidal light curves to the quiescent $B$, $V$,
$R$ and $I\/$ band data (OB), kindly provided by J. A. Orosz, using a
standard chi-squared test.  The quiescent data are much better sampled
in the $V$ and $I$ bands than the in $B$ and $R\/$ bands (there are 6-7
times as many data points in $V$ and $I$ compared to $B$ and
$R$). This has the consequence of giving more weight to the $V$ and
$I$ data points and hence produces a better fit in these passbands
than in $B$ and $R$. OB tried to compensate for this by giving each
point in the $B$ and $R$ passbands seven times the weight in their
fitting procedure than in the other passbands. However, this gives too
much weight to any outlying points the data contains as the number of
$\sim$ 25 points in each of these two passbands is not large enough
to eliminate small number statistical effects (e.g. the different $R$
magnitudes at phases 0.25 and 0.75 of the data cannot be modelled
properly). We therefore chose to give each point equal weight. A grid
search was used to find the best-fitting model as this is the most
reliable way of finding the global minimum. Since the measurement
errors are not normally distributed, it is not valid to establish
quantitative relationships between $\Delta\chi^2$ and the confidence
limits. The $\chi^2$ test therefore provides a merit function not a
maximum likelihood estimator. Because of this, the Hessian matrix
(the inverse of the covariance matrix) was not used for error
estimation, but a constant $\Delta\chi^2$ was chosen as the boundary to
define confidence regions.

The parameters of the model are highly correlated with each other. For
example, an increase in distance (which makes the object fainter) can
be compensated, at least to some degree, in several different ways:
(1) by a decrease in mass ratio, which increases the secondary's
Roche-lobe radius and hence increases its surface area; (2) a decrease
in $E(B-V)$, which reduces the extinction; or (3) an increase in
temperature, which makes the object more luminous.  The distance to
Nova Sco is relatively well determined from the kinematics of the
radio jet (Hjellming \& Rupen 1995), and we use this as an additional
constraint in the modelling,
in order to obtain tighter limits on the various model parameters. The
colour excess and temperature of the secondary can be used to estimate
the spectral type. This allows us to check whether the best-fitting
parameters are consistent with the observed spectral type of Nova Sco.
This is preferable to the previous analyses, which allowed an
arbitrary shift in the light curves, thereby losing information on the
distance and extinction. Shifting the light curves also has the effect
of making the light-curve amplitudes strongly dependent on the
proportion of disc flux. The present method avoids this problem as
altering the proportion of disc flux shifts the light curves while
also altering their amplitudes. Therefore, an overestimate of the disk
flux will be seen as a shift in the light curves toward brighter
magnitudes. If the light curves were arbitrarily offset, this would
only appear as a change in amplitude which could be compensated by a
change in inclination, mass ratio or temperature.

The treatment of the distance and the colour excess as free
parameters, and the inclusion of model atmosphere fluxes, are the only
significant differences between our model and those of OB and
vdH. Indeed, we have checked that, using the same assumptions as OB
and vdH, we obtain light-curve fits that are essentially identical to
those in these respective studies.

\section{Light-curve modelling} \label{lcmodel}
\subsection{The disc structure}
In our initial modelling we assumed a steady-state disc 
where the temperature profile follows a $-3/4$ power law (i.e.
$T_{\rm disc}\propto r^{-3/4}$). This corresponds to an optically
thick disc where each element emits like a blackbody (Shakura
\& Sunyaev 1973). In such a disc, most of the contribution to the
observed disc flux comes from the inner parts of the disc.
We systematically varied the temperature at the disc rim, but did not
find any model that produced a good fit to the observed light curves
since all models produced too much blue light in the inner parts of
the disc. The best fit was obtained for a cool accretion disc with
a rim temperature of 1000\,K, which had a $\chi^2_{\nu\rm{,min}}$
of 2.1. However, in this model the fitted distance was much closer
than the distance of $3.2 \,\pm\, 0.2$ kpc found by Hjellming \& Rupen
(1995); hence this model also had to be discarded.

We then considered two alternative types of disc model: (1) an
accretion disc with a central hole and (2) a disc with a flat
temperature profile. 

\begin{figure*}
\begin{center}
\epsfig{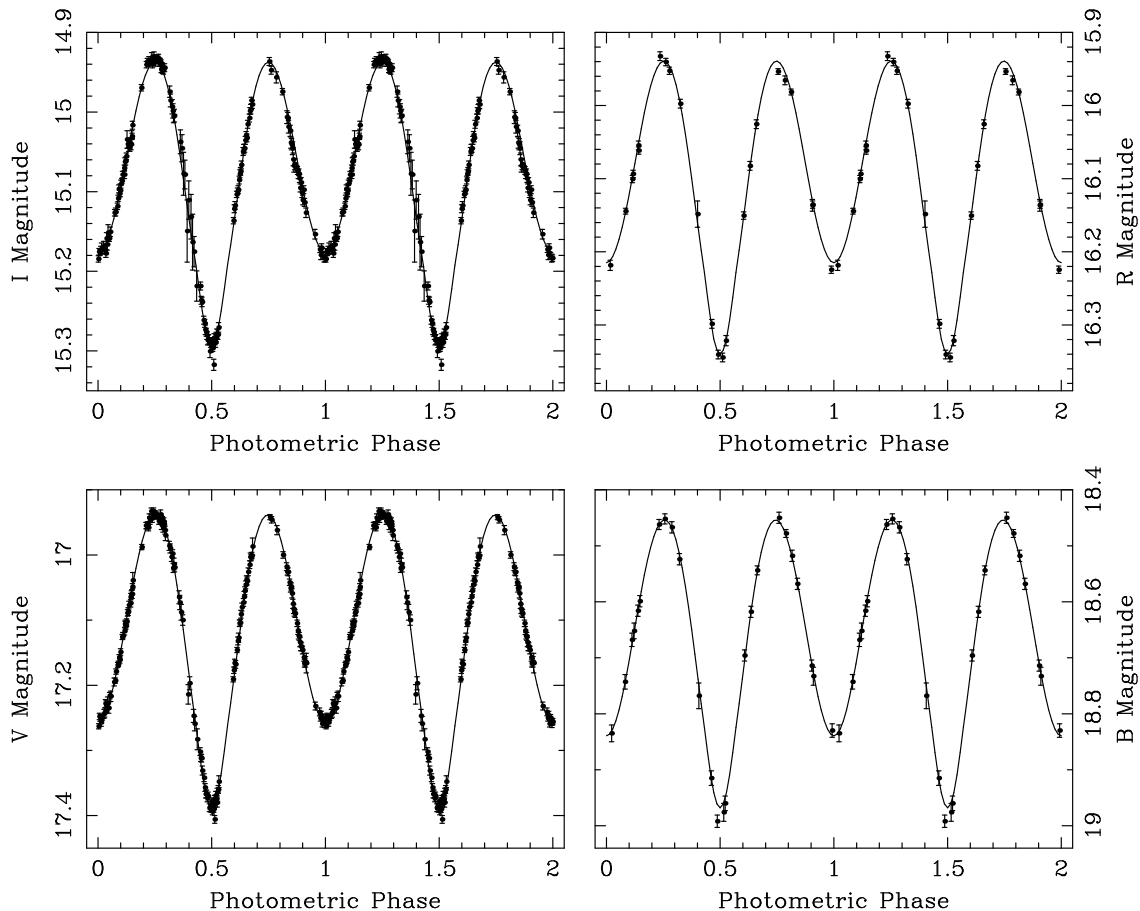}
\caption{Fits to the data for the best-fitting parameters.}
\label{bestfits}
\end{center}
\end{figure*}

A central hole in the accretion disc of GRO J1655--40 has previously
been proposed by Hameury et al.\ (1997) to explain the time delay
between the optical and the X-ray outbursts. Models for discs with
central holes have also been suggested by various authors for a variety
of systems (e.g. Meyer \& Meyer-Hofmeister 1994;
Meyer-Hofmeister \& Meyer 2000; Narayan, McClintock \& Yi 1996).  In
these models the hole is caused by evaporation of the inner parts of
the disc. Here, we assumed that the temperature at the inner edge of
the disc did not exceed 7000\,K (a higher temperature would produce
too much blue light). The 7000 K cutoff corresponds to $r =
0.067\,r_{\rm{L}}$ and $r = 0.17\,r_{\rm{L}}$ for rim temperatures
of 1000 and 2000\,K, respectively.

In cataclysmic variables (CVs), maximum entropy eclipse mapping (MEM)
has been used to determine the radial temperature profile of the discs
(Horne 1993). In outburst, the disc temperature profile is generally
found to be consistent with a $-3/4$ power law, whilst in quiescence
the profile is significantly less steep, occasionally even flat (Wood
et al.\ 1986, 1989). It has been suggested that the temperature
profile in SXTs will also be flat during quiescence (Janet Wood,
private communication).  In their study, OB let the power-law index of
the disc temperature profile be a free parameter and found a value of
$-0.12$ for their best-fitting model. Based on this finding, we
adopted a power-law index of $-0.1$ for our model with a flat
temperature profile. Since for such a profile, most of the flux comes
from the outer parts of the disc, this model is not sensitive to the
chosen inner disc radius.
 
In models with a disc rim temperature of 1000\,K, the disc does not
contribute much flux to the system. Its principle effect is in
eclipsing the secondary and thereby increasing the depth of the
minimum at phase 0.5. Therefore, the addition of a central hole did
not improve the fit. The best-fitting model for the disc with the
central hole was obtained for a rim temperature of 2000\,K. This model
has a reduced $\chi^2_{\nu\rm{,min}}$ of 1.7 for 370 degrees of
freedom. The 2000\,K disc contributes a greater proportion of the
system's flux than the cooler 1000\,K disc. The disc with the flat
temperature profile was modelled so that it produced the same amount
of flux as the 2000\,K disc with the central hole ($\sim 5$ per cent
in $V$). This corresponds to a rim temperature of 3500\,K at 0.9
$r_{\rm{L}}$. This model produced the overall best fit with a
$\chi^2_{\nu\rm{,min}}$ of 1.65.

To improve the fits further, we then used a finer grid with spacings
of 0.1 in the mass ratio, 0.1\degr\ in inclination, 0.001 in
$E(B-V)$, 0.002 kpc in distance and 25 K in polar temperature for the
two disc models. The results of the best fits are shown in
Table~\ref{bfvalues} for the two types of disc models.  The
light-curve fits for the best-fitting model with a flat disc profile
and outer disc radius of $0.8\,r_{\rm{L}}$ are shown in
Fig.~\ref{bestfits} (this model has a reduced
$\chi^2_{\nu\rm{,min}}$ of 1.614).

\begin{table*}
\caption{Best-fitting parameters for models with a flat disc
temperature profile and for a steady-state disc with an inner disc 
temperature of 7000\,K and disc rim temperature of 2000\,K.}
\label{bfvalues}
\begin{center}
\begin{tabular}{lccccccc}
\hline
Parameter & \multicolumn{7}{c|}{Model} \\
\rule[-4pt]{0pt}{12pt}
	& \multicolumn{3}{c|}{$r_{\rm{out}}$/$r_{\rm{L}}$ for
$T_{\rm{disc}} \propto r^{-0.1}$} & &
\multicolumn{3}{c|}{$r_{\rm{out}}$/$r_{\rm{L}}$ for $T_{\rm{in}}
= 7000$ K} \\
  & 0.7 & 0.8 & 0.9 & & 0.7 & 0.8 & 0.9	\\ \hline
Mass ratio & 3.4 & 3.9 & 4.0 & & 3.9 & 4.4 & 5.3  \\ 
Inclination & 69\fdg 5 & 68\fdg 4 & 67\fdg 3 & &
70\fdg 0 & 68\fdg 7 & 67\fdg 4\\ 
$T_{\rm{pole}}$ (K) & 6625 & 6525 & 6350 & & 6850 & 6750 & 6725 \\
$E(B-V)$ & 1.037 & 1.001 & 0.950 & & 1.089 & 1.067 & 1.055 \\
Distance (kpc) & 3.340 & 3.208 & 3.248 & & 3.138 & 2.998 & 2.820 \\
Minimum $\chi^2_{\nu}$ & 1.644 & 1.614 & 1.764 & & 1.697 & 1.675 & 1.673 \\
  & & & \\
$M_1$ ($\rm{M}_{\sun}$) & 5.65 & 5.35 & 5.45 & & 5.20 & 5.10 & 4.90 \\
$M_2$ ($\rm{M}_{\sun}$) & 1.65 & 1.40 & 1.35 & & 1.35 & 1.15 & 0.90 \\
$L_2$ ($\rm{L}_{\sun}$) & 25.0 & 20.5 & 18.5 & & 24.5 & 21.0 & 17.5 \\
$T_{\rm{eff}}$ (K) & 6200 & 6100 & 5925 & & 6450 & 6325 & 6325 \\ \hline
\end{tabular}
\end{center}
\end{table*}

\begin{table*}
\caption{90 per cent confidence limits for a steady-state disc with a central
hole. The inner disc temperature of 7000 K corresponds to
$r_{\rm{in}}$ of 0.17 $r_{\rm{L}}$.} \label{bfhole}
\begin{center}
\begin{tabular}{lccc}
\hline	
Parameter & \multicolumn{3}{c|}{Outer disk radius}\\
 & 0.7 & 0.8 & 0.9	\\ \hline
Mass ratio & 3.65 - 4.15 & 4.20 - 4.75 & 5.00 - 5.90 \\ 
Inclination & 69\fdg 75 - 70\fdg 15 & 68\fdg 50 - 68\fdg 90 & 67\fdg
20 - 67\fdg 65	\\ 
$T_{\rm{pole}}$ (K) & 6720 - 6935 & 6670 - 6875 & 6610 - 6785 \\
$E(B-V)$ & 1.057 - 1.117 & 1.039 - 1.103 & 1.030 - 1.074 \\
Distance (kpc) & 2.995 - 3.296 & 2.890 - 3.096 & 2.685 - 2.912 \\
& & & \\
$M_1$ ($\rm{M}_{\sun}$) & 5.05 - 5.35 & 4.95 - 5.20 & 4.70 - 5.00 \\
$M_2$ ($\rm{M}_{\sun}$) & 1.20 - 1.45 & 1.05 - 1.25 & 0.80 - 1.00 \\ 
$L_2$ ($\rm{L}_{\sun}$) & 22.5 - 26.5 & 19.5 - 22.5 & 15.5 - 19.0 \\ 
$T_{\rm{eff}}$ (K) & 6250 - 6500 & 6225 - 6425 & 6150 - 6350 \\ \hline
\end{tabular}
\end{center}
\end{table*}

\begin{table*}
\caption{90 per cent confidence limits for a disc with a flat
temperature profile.} \label{bfflat}
\begin{center}
\begin{tabular}{lccc}
\hline	
Parameter & \multicolumn{3}{c|}{Outer disk radius}\\
 & 0.7 & 0.8 & 0.9	\\ \hline
Mass ratio & 3.30 - 3.60 & 3.55 - 4.00 & 3.65 - 4.50 \\ 
Inclination & 69\fdg 35 - 69\fdg 75 & 68\fdg 25 - 68\fdg 60 & 67\fdg
10 - 67\fdg 60	\\ 
$T_{\rm{pole}}$ (K) & 6550 - 6745 & 6380 - 6570 & 6210 - 6410 \\
$E(B-V)$ & 1.017 - 1.067 & 0.970 - 1.019 & 0.917 - 0.962 \\
Distance (kpc) & 3.167 - 3.496 & 3.116 - 3.398 & 3.106 - 3.352 \\
 & & & \\
$M_1$ ($\rm{M}_{\sun}$) & 5.35 - 5.70 & 5.30 - 5.60 & 5.15 - 5.70 \\
$M_2$ ($\rm{M}_{\sun}$) & 1.45 - 1.80 & 1.30 - 1.60 & 1.10 - 1.60 \\ 
$L_2$ ($\rm{L}_{\sun}$) & 22.5 - 27.0 & 19.0 - 23.0 & 15.0 -
20.5 \\ 
$T_{\rm{eff}}$ (K) & 6100 - 6300 & 5950 - 6150 & 5750 - 6025 \\ \hline
\end{tabular}
\end{center}
\end{table*}

\subsection{Confidence limits for the system parameters}
The fitting procedure has two sources of errors: one is the general
statistical error and the other is caused by the finite size of the
grid.  Ideally, an iterative process would be used to find the global
minimum at each value of a parameter. However, we found that there
were a large number of minima and that an iterative process was not
guaranteed to find the global minimum. To estimate the effect of the
finite grid size, we used the following procedure. At the minimum we
estimated the effect of the finite grid size on each parameter in turn
by fixing that parameter to the value it has at the minimum and by
varying the other parameters systematically. For each parameter we
calculated the increase in $\chi^2$ corresponding to a value half a
grid spacing away. The largest increase in the value of $\chi^2$ was
then taken as the uncertainty, $\Delta\chi^2$, in $\chi^2$ due to the
grid resolution for the particular parameter kept fixed. The 90 per
cent confidence limits generally generate an increase in $\chi^2$ of
2.71 (Avni 1976). We added this to $\Delta\chi^2$ and used the
resulting value to define the 90 per cent confidence region and to
obtain the confidence limits for each individual parameter.

The masses of the two binary component masses are not parameters that
are fitted directly, but are determined from the mass function
(kept fixed) and the values of the inclination and mass ratio at
a particular grid-point in the five-dimensional model parameter space.
To determine the uncertainty in $\chi^2$ due to the finite grid size,
we used a similar method as above, except that this time both the 
mass ratio and the inclination of the system were held fixed at their
values at the minimum and the remaining three parameters were varied
to find $\Delta\chi^2$. The 90 per cent confidence limits for a
quantity depending on two parameters generates an increase in $\chi^2$
of 4.61 (Avni 1976). This was added to $\Delta\chi^2$ and the result
was used to define the confidence regions for the component
masses. Similarly, to calculate the confidence limits for the
luminosity ($L_2$) and effective temperature ($T_{\rm{eff}}$) of the secondary, the same method
was used, except that these quantities depend on three parameters: the
mass ratio, the inclination and the polar temperature. The luminosity
was calculated by summing $\sigma T^4$ over the surface, and the
effective temperature by dividing the luminosity by the surface area.
The effect of the finite size of the grid in the distance and colour
excess was then found and added to the increase in $\chi^2$,
corresponding to the 90 per cent confidence limits for a quantity
depending on three parameters (6.25, Avni 1976).

The 90 per cent confidence limits for the steady-state disc models
with a central hole with an inner edge disc temperature of 7000 K are
shown in Table~\ref{bfhole}. This inner edge disc temperature
corresponds to $r_{\rm{in}} = 0.17\,r_{\rm{L}}$. Table~\ref{bfhole}
shows that, as the outer disc radius is increased in this model, the
mass ratio increases and the distance decreases while the secondary
temperature and luminosity remain the same. The secondary temperature
and luminosity are consistent with the observed $B-V$, whilst only the
outer disc radius of $0.7\,r_{\rm{L}}$ has a distance which is
consistent with that of Hjellming \& Rupen (1995). (The model with an
outer disc radius of $0.8\,r_{\rm{L}}$ is marginally consistent.)

The 90 per cent confidence limits for the parameters and the
derived quantities are shown in Table~\ref{bfflat} for the disc model
with a $-0.1$ temperature profile. Table~\ref{bfflat} shows that, as
the outer disc radius is increased in this model, the temperature of
the secondary decreases, as does the colour excess. This is consistent
as a lower temperature implies a later spectral type which has a
larger $(B-V)_0$. The distances are similar to the measurement of
Hjellming \& Rupen (1995) for all values of the outer disc radius.

Comparison of the various models in Tables~\ref{bfhole} and
\ref{bfflat} shows that the variation in the parameters from model to
model is larger than the statistical errors for each individual model
(as given in the tables) and that the systematic errors due to the
uncertainties in the modelling are the dominant sources of error. We
can use this variation of parameters as an {\it indication} of the
systematic errors, although we need to emphasize that this is really
only a lower limit, since we only examined a limited number of
models. Using the distance measurement by Hjellming \& Rupen (1995) as
an additional discriminant, we obtain the best-guess estimates for the
parameters given in Table~\ref{bestfit}  with uncertainties that
include both the statistical errors and an estimate of the systematic
errors.

\begin{table}
\begin{center}
\caption{Best-fitting model parameters for Nova Sco} \label{bestfit}
\begin{tabular}{lc}\\
\noalign{\vspace{-10pt}}
\hline 
\noalign{\vspace{1pt}}
Mass ratio & $3.9 \pm 0.6$ \\
Inclination & $68\fdg 65 \pm 1\fdg 5$ \\
$T_{\rm{pole}}$ & $6575 \pm 375\,$K \\
$M_1$ & $5.40 \pm 0.30\, \rm{M}_{\sun}$ \\
$M_2$ & $1.45 \pm 0.35\, \rm{M}_{\sun}$ \\
$L_2$ & $21.0 \pm 6.0\, \rm{L}_{\sun}$ \\
$T_{\rm{eff}}$ & $6150 \pm 350\,$K \\
$E(B-V)$ & $1.0 \pm 0.1$ \\
\hline
\end{tabular}
\end{center}
\end{table}

\subsection{Checking for self-consistency}
\subsubsection{The colour and spectral type of the secondary}
To determine the intrinsic $(B-V)_0$ of the secondary, the
contribution of the disc to the overall $B-V$ has to be
subtracted. This was done by calculating the $B-V$ for the
best-fitting models when the disc flux was not included. This resulted
in a decrease in $B-V$ of 0.01\,--\,0.02. The $E(B-V)$ of the
best-fitting models then leads to a $(B-V)_0$ of 0.42\,--\,0.61 for
the secondary, which corresponds to a spectral-type range of
F4\,--\,G0 (Fitzgerald 1970). This is a somewhat later spectral type
than found in previous estimates but is consistent with the analysis
using quiescent data (S99). The spectral type can also be determined
from the effective temperature ($6150\,\pm\, 350\,$K). This
temperature range corresponds to the spectral-type range F5\,--\,G2
(Strai$\check{\rm{z}}$ys \& Kuriliene 1981). The two measurements are
consistent with each other, implying a spectral type for the secondary
of F5\,--\,G0. We note, however, that due to the highly anomalous chemical
composition of the secondary (Israelian et al.\ 1999), it is not clear
how well these standard relations apply.

In our model we have assumed that the temperature distribution of the
secondary star is described by a gravity-darkening law appropriate for
a radiative atmosphere (von Zeipel 1924), for which the local
temperature is proportional to the local value of gravity to the 0.25
power. If the atmosphere were convective, we would expect a 
gravity-darkening coefficient of 0.08 (Lucy 1967). The deduced spectral type
of the secondary (F5\,--\,G0) is on the boundary between hot stars
with radiative atmospheres and cool stars with convective
atmospheres. If the atmosphere were convective (with a coefficient of
0.08), then the temperature variation over the surface would be
smaller than in the radiative case, and the system would require a
higher inclination to give the same light-curve amplitude. However, a
larger inclination would produce deeper, sharper eclipses which are
not observed. This suggests that the steeper gravity-darkening
coefficient for radiative atmospheres is indeed the most appropriate
to use. Near the L1 point, the star's temperature ($\sim\! 4000\,$K)
is cooler than the polar temperature because of the lower surface
gravity (2.5 vs.\ 3.5 dex). This implies that the L1 point may be
convective rather than radiative. However, this would only affect the
light curve near phase 0.5, when the accretion disc is in front of the
secondary. Thus, any difference could be represented by a variation in
the amount of eclipsing, i.e. the size of the disc. As we have
considered three different disc sizes, we can be confident that this
effect does not significantly add to the uncertainties.

Unlike the previous studies, we considered a large range of
temperatures and colour excesses (and hence luminosities), both of
which have been consistent with the spectral type. We have, however,
assumed a mean Galactic extinction curve in calculating the
extinction.  The model is very dependent on the extinction curve as
this affects the relative offsets of the light curves in different
passbands. If the extinction curve were incorrect, then the fitting
procedure would compensate for this by choosing different values for
the distance, temperature and $E(B-V)$. Our temperature and $E(B-V)$
are in good agreement with the `observed' spectral type, which implies
that the actual extinction cannot significantly deviate from the mean
Galactic curve. The greatest variations in Galactic extinction occur
in the UV (Fitzpatrick 1999). Since our fits do not rely on UV data,
we suspect that deviations from the Galactic mean are unlikely to be
important. The alternative to using the mean Galactic extinction curve
would have been to shift the light curves in the modelling. However, as
discussed previously, shifting the light curves arbitrarily is not
desirable (since this leads to the loss of information) and so using
the mean Galactic extinction curve is preferable, especially as the
modelling proves to be fully self-consistent.

\subsubsection{Modelling the disc} \label{discmodel}

Our disc model is rather simple. For example, we assumed a flat
cylindrical disc; this is unrealistic as the disc will almost
certainly be more complicated with well defined structure. Possible
evidence of structure can be seen in the $I$ band data near phase 0.85
where there is a systematic offset between the data points and the
light curve. The data points are dimmer with flux variations of up to
2.5 per cent. Even though the disc in our model contributes only a
small fraction of the total flux ($\sim\! 5\,$per cent in $V$), the model is
sensitive to its contribution (as can be seen from the comparison of
the best-fitting models for different disc structures in
Tables~\ref{bfhole} and \ref{bfflat}).  In our modelling, we varied
the size, temperature and power-law index of the disc. The size of the
disc determines the depth of the grazing eclipse and hence the
inclination. Since we varied these key disc parameters over a wide
range of plausible values, we expect our estimates of the model
parameters (e.g. the inclination and the component masses) and their
uncertainties to be reasonably realistic.

The proportion of flux, the cool accretion discs in our models
contribute to the total flux, increases with wavelength i.e. the
minimum contamination is in $B$. This is different from models for the
SXT A0620--00 (V616 Mon) where the disc contribution appears to
decrease from $43 \pm 6$ per cent in $V$ (Oke 1977; McClintock \&
Remillard 1986) to less than 27 per cent in the infrared ( Shahbaz,
Bandyopadhyay \& Charles 1999). This shows that even in A0620--00,
which has a much shorter orbital period (7.75 hr; McClintock \&
Remillard 1986) and has been in quiescence for much longer, the disc
contributes appreciably to the total flux of the system.

On the other hand, Greene et al.\ (2001) have justified the use of a
model without a disc by claiming that an accretion-disc model does not
fit the data well and that the flux contribution from it will be
negligible. The disc model they consider, however, contributes a large
proportion of the $K$ flux, as evidenced by the eclipses in their
accretion-disc model at phase 0.0 as opposed to phase 0.5 for eclipses
of the secondary. A model with a smaller proportion of $K$ flux would
have a larger amplitude than their accretion-disc model. This would
better fit the data, decreasing the requirement for a model without an
accretion disc. In the next section we will show that models without
an accretion disc affect other model parameters, in particular the inferred
distance, and that no fully self-consistent models can be obtained
without a disc.

\subsubsection{J and K magnitudes} \label{greenecomp}

Our model can also be used to make predictions for other
passbands. Using our best-fitting models, we calculated the expected
$J$ and $K$ band light curves. For the $J$ and $K$ bands we find
mean magnitudes of 13.8 and 13.0, respectively, with the disc
contributing between 10 and 20 per cent of the total flux depending on
the model. Greene et al.\ (2001) have recently 
presented $BVIJK$ photometry of GRO J1655--40 in quiescence. They find
mean $J$ and $K$ magnitudes of 13.85 and 13.25, respectively. Our $J$
band prediction is in good agreement with their photometry apart from
the depth of the minima, which is somewhat too low (possibly because
the inclination is slightly too low). However, the general agreement
in the $J$ band provides some direct confirmation that the luminosity
and colour excess found in our modelling are accurate. The amplitude
of our $K$ band light curve is also in good agreement with their
photometry. We find a slightly larger amplitude than their
best-fitting model but entirely consistent with their data. The reason
for the $K\/$ band mean magnitude discrepancy is not clear, although
we have only calculated the best-fitting models and so do not know
what range of $K$ magnitudes is represented by our range of
parameters.

Johnson (1966) provides intrinsic infrared colours for luminosity
classes III and V. We may compare these to the colour implied by the
photometry of Greene et al.\ (2001). Their $J-K$ colour is 0.6.
Assuming the standard $J$ and $K$ extinction coefficients of
Fitzpatrick (1999), i.e. $(J-K)_0 = (J-K) - 0.5 E(B-V)$, and our
$E(B-V)$ of $1.0\,\pm\,0.1$, we obtain a $(J-K)_0$ of $0.1\,
\pm\,0.05$. For luminosity class III, Johnson (1966) only lists the
infrared colours for spectral type G5 and later. For luminosity class
V, Johnson (1966) gives $(J-K)_0$ colours of 0.28 and 0.32 for spectral type
F5 and G0, respectively -- a $(J-K)_0$ colour of 0.1 would correspond
to an A star. The $(J-K)_0$ colour appropriate for a F5 or G0 star
would require a $J-K$ of 0.8, similar to the value in our model. Hence
there appears to be a deficit of $K$ flux in the system, the cause of
which is not clear.

We can easily check whether a similar discrepancy exists in the
modelling of Greene et al.\ (2001). Using the parameters found in 
their modelling, we calculated how
much their light curves have been shifted in each passband and
determined the {\it implied} $E(B-V)$ and distance in their model. Using the
shifts in all five bands, we find an $E(B-V)$ of $1.041\,\pm\,0.022$
and a distance of $4.008\,\pm\,0.045\,$kpc. For this fit, however, the
mean magnitudes in the different passbands are not in reasonable
agreement with the data, mainly because of the inclusion of the $K$
band. If we repeat the analysis without using the $K$ band, we find an
$E(B-V)$ of $1.076\,\pm\,0.003$ and a distance of
$3.805\,\pm\,0.013\,$kpc. This provides good agreement with the mean
$B$, $V$, $I$ and $J$ magnitudes but produces a mean $K$ magnitude of
13.06. Hence the discrepancy in the $K$ band is present in their
modelling as well. We also note that the implied distance in their
model is substantially larger than that found from the kinematics of
the radio jet. The reason is that their secondary is too massive and
hence too luminous, requiring the system to be more distant in order
to have the correct visual magnitude.

We ran our model without an accretion disc to see if a self-consistent
model could be obtained in this way when fitting to the $B$, $V$, $R$
and $I$ data. We found a
$\chi^2_{\nu,\rm{min}}$ of 3.120, substantially larger than the models
with the accretion discs. Table~\ref{nodisc} shows the best-fitting
parameters for this case and the best-fitting values of Greene et al.\
(2001) who assumed a fixed effective temperature and who did not
consider the distance or colour excess in calculating their light
curves (the values shown are those implied by their model as described
above).

It is clear that the secondary in our model is similar to the
secondary in their model, with a similar mass, temperature and
luminosity. We find, however, a higher inclination. This is due to the
higher temperature of the secondary which is correlated with the
inclination. We note, however, that the secondary is similar and that
in eclipsing systems there is a strong constraint on the inclination
due to the depth of the eclipse. Hence, we can be satisfied that both the
inclination and the other system parameters are accurate in the
previous models. The lower mass ratio is clearly a consequence of
neglecting the accretion disc.

Due to the large $\chi^2_{\nu}$ of the model without an accretion
disc, we therefore conclude that an accretion is required for 
any self-consistent model of Nova Sco.

\begin{table}
\begin{center}
\caption{Best-fitting parameters for a self-consistent model without a
disc compared to the results of Greene at al.\ (2001).} \label{nodisc}
\begin{minipage}{8cm}
\begin{tabular}{lcc} \\
\hline
Parameter & \multicolumn{2}{c}{Model} \\
 & Self-consistent & Greene et al. (2001) \\ \hline
Mass ratio & 2.4 & $2.6 \pm 0.3$ \\
Inclination & 77\fdg 7 & $70\fdg 2 \pm 1\fdg 9$ \\
$E(B-V)$ & 1.127 & 1.076\footnotemark  \\
Distance (kpc) & 3.712 & 3.805\addtocounter{footnote}{-1}\footnotemark \\
$T_{\rm{pole}}$ (K) & 6950 & 6768 \\
Minimum $\chi^2_{\nu}$ & 3.118 & 1.612 \\
 & & \\
$M_1$ ($\rm{M}_{\sun}$) & 5.9 & $6.3 \pm 0.5$\\
$M_2$ ($\rm{M}_{\sun}$) & 2.4 & $2.4 \pm 0.4$ \\ 
$L_2$ ($\rm{L}_{\sun}$) & 40.1 & 31.9\,--\,40.6 \\ 
$T_{\rm{eff}}$ (K) & 6500 & 6336 (Fixed) \\ \hline
\end{tabular}

\addtocounter{footnote}{-1}
\footnotemark
\vspace{-26pt}
\footnotetext{Implied from their data (see Section~\ref{greenecomp}).}
\end{minipage}
\end{center}
\end{table}

\section{The variation in surface temperature and implications for the 
determination of the spectral type} \label{sectemp}
As a result of gravity darkening, there is a large variation in
temperature across the surface of the secondary, and hence an observer
will see a variation of the secondary's temperature as a function of
orbital phase. To determine the magnitude of this effect, a
flux-weighted average temperature and its standard deviation were
calculated at each orbital phase. For the flux-weighting we used the
$R$ band which has a wavelength range similar to that of the spectra
previously used for the determination of the spectral type
(6350\,--\,6750\AA).

\begin{figure}
\begin{center}
\epsfig{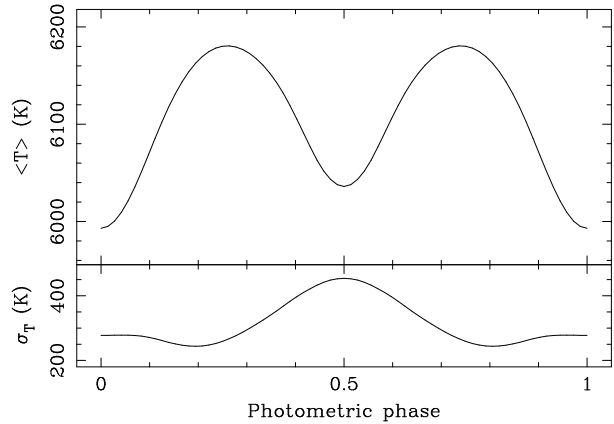}
\caption{The variation in the flux-weighted average temperature and
its standard deviation for the best-fitting model as a function of
phase.} \label{tmean}
\end{center}
\end{figure}

Fig.~\ref{tmean} shows the variation in average temperature and the
standard deviation in the flux-weighted temperature distribution with
phase for the best-fitting model parameters of
Section~\ref{lcmodel}. It demonstrates that the average `observed'
temperature of the secondary is substantially lower ($\sim\!450\,$K)
than the temperature at the poles where the gravity is largest. There
is also a clear variation in average temperature with phase ($\sim\!
200\,$K). This variation would be large enough to change the observed
spectral sub-type with phase if there was not such a large range of
temperatures across the observable part of the secondary at each
phase. The total range within  one standard deviation is
5600\,--\,6500\,K. This range corresponds to stars of spectral
sub-type F5\,--\,G5 (Strai$\check{\rm{z}}$ys \& Kuriliene 1981). The
effect this variation in temperature has on the observed line profiles
is unknown and is worth further investigation. We conclude, however,
that it is difficult to determine the spectral type to better than a
few sub-types due to this large variation in surface temperature.

\section{Discussion} \label{discuss}

\subsection{Comparison to previous studies}

The accretion-disc model with a flat temperature profile provides the
best-fitting model. This result is consistent with disc structures in
CVs during quiescence (based on MEM), which appear to have temperature
profiles that are much flatter than those expected for steady-state
discs. VdH found that, for a steady-state disc, there were no
reasonable solutions with rim temperatures greater than 1000\,K. This
is consistent with our results as the models with a 2000\,K disc do
not provide good fits to the data unless the inner disc is removed. OB
allowed the power-law exponent for the temperature profile, the rim
temperature and the size of the disc to be free parameters. They
obtained a best-fitting power-law exponent of $-0.12\,\pm\,0.01$, a rim
temperature of $4317\,\pm\,75\,$K and an outer disc radius of
$0.747\,r_{\rm{L}}$. With an exponent of $-0.1$, we fit the data with
a disc rim temperature of 3500\,K at $0.9\,r_{\rm{L}}$, which at the
radius $0.747\,r_{\rm{L}}$ has a temperature of 3580\,K. Hence the disc
in this model is 750\,K cooler than that found by OB. The accretion
disc model with a central hole contributes predominantly red light as
it has a low average temperature, similar to the disc with the flat
temperature profile. Hence we can conclude that the disc in GRO
J1655--40, when it is in quiescence, is predominantly red and
cool. Although the flatter temperature profile is preferred in the
modelling, the actual disc will almost certainly be more complicated.

OB only considered one accretion disc structure in their model,
although their accretion-disc parameters were free to vary. Our analysis
shows that the system parameters vary with the accretion-disc
model. Hence, a range of accretion-disc models need to be considered
in order to avoid underestimating the systematic errors. Relying on
the minimization of $\chi^2$ is insufficient as the
modelling is not sufficiently accurate to enable a reliable
determination of the precise accretion-disc structure. This is
because, as mentioned earlier, the actual accretion-disc will be more
complicated than the simple model assumed in these analyses.

Indeed, from our analysis, it is clear that a large range of mass
ratios and hence masses will fit the light-curve amplitudes. These
amplitudes are strongly dependent on the accretion-disc model and the
proportion of flux for each model in each band. 
GRO J1655--40 has a secondary of earlier spectral type and
hence higher luminosity than most other SXTs. This means that 
accretion-disc contamination is less important than in other systems,
although it still strongly affects the inferred system paramaters.
We therefore used the independent distance determination from the kinematics
of the radio jet (Hjellming \& Rupen 1995), as used previously
to model the jets in SS\,433 (Hjellming \& Johnston
1988) to further constrain the allowed  disc models. 
There are, however, a number of possible systematic errors in
this measurement. The inclination of the jet axis of $85\degr
\,\pm\, 2\degr$ is significantly different from the system
inclination of $68\fdg 65 \,\pm\, 1\fdg 5$, implying a more
complicated geometry than the simple model they use. This model also 
does not describe all the structure observed in the jet, and there are
deviations from simple linear expansion which they assume. In addition
there is motion over the length of an observation which would result
in smearing. On the other hand, they chose a beam-width larger than 
the proper motion during each six-hour VLBA
observation to minimize this effect. These observations did not have
absolute positional information resulting in the central source being
used for reference. This is undesirable as it makes the alignment
of the different images in their analysis a free parameter. 
If this distance measurement were incorrect, this could somewhat alter our
best-fitting parameters. Distance estimates by other
authors are, however, consistent with their determination: ${\sim \! 3}\,$kpc
(Bailyn et al.\ 1995a); ${\sim \! 3}\,$kpc (Greiner, Predehl \& Pohl 1995);
$3.5\,$kpc (McKay \& Kesteven 1994) and $3\,$--$\,5\,$kpc (Tingay et
al.\ 1995). Hence, using their distance determination of
$3.2\,\pm\,0.2\,$kpc as additional constraint is reasonable.
	
A lower value for the colour excess than used previously is needed to
provide good fits to the data in all passbands simultaneously.  We
found an $E(B-V)$ of $1.0\,\pm\,0.1$ (90 per cent confidence). This value is
somewhat, but not dramatically lower than various previous estimates
(1.15, Bailyn et al.\ 1995a; 1.3, Horne et al.\ 1996; $1.2\,\pm\,0.1$,
Hynes et al.\ 1998, 1$\sigma$ limits). This $E(B-V)$ along with the
effective temperature implies a spectral type range for the secondary
of F5\,--\,G0, consistent with previous estimates (S99). VdH
used a value for the colour excess of 1.3 in their study 
and hence had to assume a significantly larger luminosity of the
secondary (31\,--\,54\,L$_{\sun}$ as compared to
$21.0\,\pm\,6.0\,$L$_{\sun}$ found here). This resulted in significantly
larger masses for both the black hole and the secondary.
A lower luminosity, which alters GRO J1655--40's position in the HR diagram, 
along with the lower masses implied by our model has implications for
the interpretation of the system's evolutionary state (see
Section~\ref{evolsec}).

Mass estimates for the components have also been obtained by Buxton \&
Vennes (2001) and Wagoner, Silbergleit \& Ortega-Rodr\'\i guez
(2001). Buxton \& Vennes (2001) fitted model spectra to the observed
quiescent spectrum and measured the rotational broadening in order to
calculate a mass ratio of 2.56\,--\,4.35 using the radial velocity
amplitude of S99 (see Section~\ref{vrot}). Combined with the inclination of
vdH the masses are $7.91 \,\pm\, 3.79\,$M$_{\sun}$ and $2.76 \,\pm\,
1.79\,$M$_{\sun}$ for the primary and secondary respectively. Wagoner
et al.\ (2001) have modelled the quasi-periodic X-ray oscillations and
find a primary mass of $5.9 \,\pm\, 1.0\,$M$_{\sun}$. Both of these
measurements are consistent with our values.

\subsection{Rotational broadening and the mass ratio} \label{vrot}

Our mass ratio of $3.9\,\pm\,0.6$ is larger than that found in the study
of OB (2.60\,--\,3.45, 3$\sigma$ limits) but is in agreement with the
upper range obtained by vdH (2.43\,--\,3.99, 3$\sigma$).  A larger mass
ratio implies a smaller and hence less luminous secondary, as required
by our lower value for the colour excess.  S99 found a mass ratio of
2.29\,--\,3.06 (95 per cent confidence) based on their estimate of the
rotational broadening of the spectral lines.  They assumed that the
secondary was spherical and that all the observed broadening was a
consequence of rotation. Fitting a broadening profile to template star
spectra and minimizing the residuals, they found a rotational
velocity, $v\sin i$, of 82.9\,--\,94.9\,km\,s$^{-1}$. Our mass ratio
range corresponds to a $v\sin i$ of 68.3\,--\,79.1\,km\,s$^{-1}$,
using their broadening to mass ratio relation. In their determination
of the average spectrum, S99 did not correct for the orbital smearing
of the spectra due to the length of the exposure time of the
observations.  Most of their spectra were taken near the quadrature
phases when orbital smearing would be at its greatest. This results in the
apparent broadening of the lines by 7\,km\,s$^{-1}$ (at the orbital
phase when the spectra were taken).  Since this broadening should be
added linearly to the rotational broadening (rather than in
quadrature), it introduces an additional, but spurious rotational
broadening of 3.5\,km\,s$^{-1}$ (about half of the change of the radial
velocity during the observations).

S99 took the radius of a spherical star in their model to be the
effective Roche-lobe radius. In reality, a Roche-lobe filling object
can be described by an ellipsoid to first approximation.  This implies
that the light contributing to a spectral line will come from a larger
range of radii (and hence velocities) than for a spherical star of the
same volume and that the assumption of sphericity will lead to an
overestimate of the rotational velocity.  The dependence of $v\sin i$
on the mass ratio for a realistic model containing a Roche-lobe
filling star is consequently non-trivial and requires detailed
modelling. Orosz \& Hauschildt (2000) have investigated the difference
in rotational broadening kernels between Roche-lobe filling models and
analytical models. They found that, for a Roche-lobe filling star, the
broadening kernel changes with phase and is significantly different
near the quadrature phases where it is wider and asymmetric. The
calculation using the analytical kernel then provides an upper limit
to $v\sin i$.  Since an upper limit on the rotational broadening
provides a lower limit to the mass ratio, we conclude that our estimates
are likely to be consistent with the findings of S99, once all of these 
effects are properly taken into account.

\begin{figure*}
\begin{center}
\epsfig{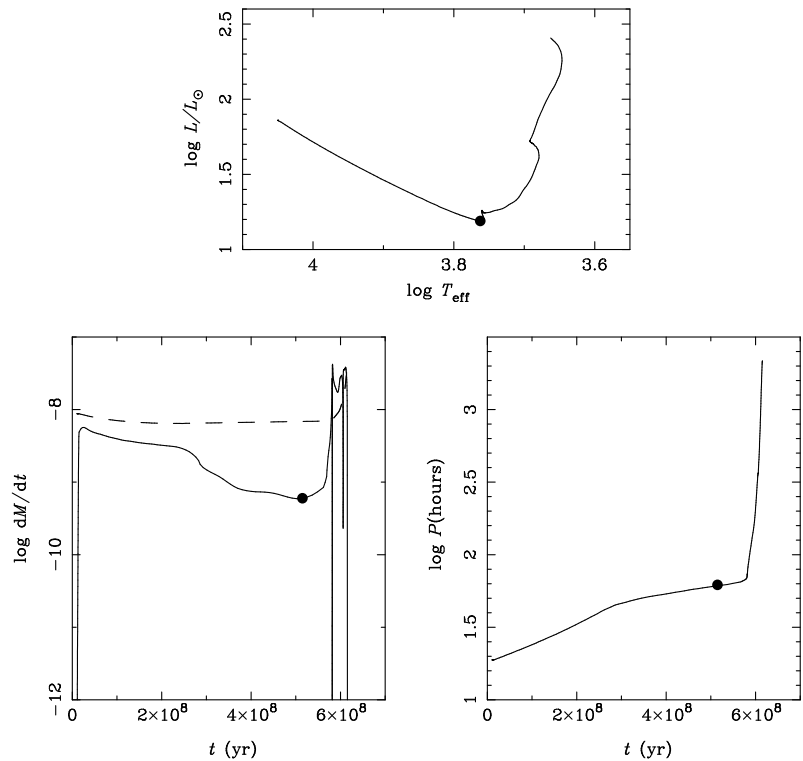}
\caption{Binary evolution model for Nova Sco (conservative 
case A mass transfer with convective overshooting of $0.3$ pressure 
scale heights). The individual panels show the Hertzsprung-Russell
diagram, the  mass-transfer rate and orbital period since the
beginning of mass transfer. The evolution after mass transfer has
ceased is not shown. The initial masses of the black hole and
secondary are 4.1 and  2.5$\,\rm{M}_{\sun}$, respectively, the
present masses are 5.3 and  1.3$\,\rm{M}_{\sun}$. The secondary is
close to exhausting hydrogen in its core. The dots indicate the values
at the orbital period of Nova Sco ($2.6\,$d). The dashed curve in the
$\dot{M}$ panel shows the critical mass-transfer rate below which
transient behaviour is expected (from King, Kolb, \& Szuskiewicz
1997).} \label{evol}
\end{center}
\end{figure*}

Greene et al.\ (2001) calculated a numerical broadening kernel
for photometric phase zero, where the Israelian et al.\ (1999)
measurements were taken. They found almost no systematic biases in
comparison to the analytical kernel, justifying the use of
$v\sin i$ of $93\,\pm\,3\,$km\,s$^{-1}$ in their model fitting. Buxton
\& Vennes (2001) have calculated $v\sin i$ by fitting model spectra to
the observed spectrum taken during quiescence and found a $v\sin i$ of $80
\,\pm\, 10$\,km\,s$^{-1}$ as well as $T_{\rm{eff}} = 6500\,\pm\,50\,$K
and [Fe/H]\,=\,$-$0.25\,--\,0.00. This is consistent with our
calculation as well as with previous measurements. Their Fe abundance
agrees with that of Israelian et al.\ (1999) who found
[Fe/H]\,=\,0.1\,$\pm$\,0.2 and larger overabundances in $\alpha-$process
elements.
Recently, Bleach et al.\ (2000) have demonstrated the unreliability of using
the rotational broadening
$v\sin i$ for the calculation of system parameters in the CV EG
UMa. They fitted broadened spectra to individual lines and found a
large range of values of $v\sin i$ of 19\,--\,59\,kms$^{-1}$, with
typical error on each value of $\sim 10\,$kms$^{-1}$. As the
velocity semi-amplitude of both components in this system is
measurable, they predict a $v\sin i$ of 28.6\,kms$^{-1}$. The reason
for this large variation is unknown, but is not due to irradiation and may
be a result of anomalous composition (for further discussion see Wood
et al.\ 2001).

\subsection{The evolutionary state of GRO J1655--40} \label{evolsec}

Modelling the binary evolution of GRO J1655--40 with a present
secondary mass of 2.35\,M$_{\sun}$ (OB) leads to an average (secular)
mass-transfer rate that tend to be much larger than is consistent with its
X-ray transient behaviour (van Paradijs 1996; King, Kolb \&
Szuskiewicz 1997).  Kolb et al.\ (1997) and Kolb (1998) showed that
this problem could be solved if the secondary were in a special
position in the Hertzsprung gap; this, however, required a somewhat
cooler secondary than is consistent with its spectral type. Reg\H os,
Tout \& Wickramasinghe (1998) suggested as an alternative solution
that the secondary is still on the main sequence. This also requires
some fine-tuning and significant convective overshooting (but is
consistent with recent findings; Pols et al.\ 1997).

The lower mass found in this re-analysis largely removes this problem,
irrespective of whether the secondary is still on the main sequence or
in the Hertzsprung gap. To illustrate this we present a case A binary
calculation (similar to the model of Reg\H os et al.\ 1998) in
Fig.~\ref{evol} (for a detailed description of the binary evolution code see
Podsiadlowski, Rappaport \& Pfahl 2001a). The dashed curve in the
panel showing the mass-transfer rate gives the critical mass-transfer
rate below which transient behaviour is expected (from King et al.\
1997). The mass-transfer rate is well below the critical rate until
the end of the main sequence. We have also performed some early case B
binary calculations (i.e. where the secondary has just evolved off the 
main sequence) and also find that, with the lower mass of the secondary,
the present mass-transfer rate is below the critical rate for
transient behaviour, although case B models generally do not
fit as well as late case A models.

In our case A model, the initial masses, i.e. immediately after the 
supernova in which the black hole was formed, were 4.1 and 2.5\,M$_{\sun}$
for the black hole and secondary, respectively. These masses
are very similar to the post-supernova masses required to explain
the pollution of the secondary with $\alpha$-process material
(Israelian et al.\ 1999) that was ejected in the supernova 
(Podsiadlowski et al.\ 2001b)\footnote{We note that,
in order to explain its anomalous composition, the secondary had
to capture some $0.25\,$M$_{\sun}$ of material mainly composed
of heavy elements, which was then mixed completely with the secondary
(Podsiadlowski et al.\ 2001b). This must have dramatically changed
the composition (in particular the metallicity) of the secondary, 
an effect that was not included in our binary calculations.}.

From all of this a complete picture starts to emerge for the
evolutionary history of Nova Sco.  The pollution of the secondary with
supernova material proves that the black hole formed in a supernova
(or a hypernova).  Some material that was produced in the
supernova had to be captured by the secondary; this requires both
fallback of material and mixing in the supernova ejecta (Podsiadlowski
et al.\ 2001b). In a typical model, the orbital period after the
binary has re-circularized after the supernova is about 20\,hr (see
the mixing models in table~3 of Podsiadlowski et al.\ 2001b), very
close to the period needed for a late case A scenario as shown in
Fig.~\ref{evol}. Since then the secondary has transferred some 1\,M$_{\sun}$,
increasing the orbital period to its present value.  Self-consistent
models also require that the black hole received a significant kick at
birth in order to produce the high observed system space velocity,
consistent with the fallback suggestion by Brandt, Podsiadlowski \&
Sigurdsson (1995).

\section{Conclusions}

We have obtained self-consistent fits to the ellipsoidal light curves
of GRO J1655--40 for the $B$, $V$, $R$ and $I$ passbands
simultaneously without making {\it a priori} assumptions about the
distance and the colour excess. Using the distance estimate of
$3.2\,\pm\,0.2\,$kpc, based on the kinematics of the radio jet, as
additional constraint, we find that our $E(B-V)$ of $1.0\,\pm\,0.1$
along with our effective temperature of the secondary of
$6150\,\pm\,350\,$K corresponds to a spectral type range for the
secondary of F5\,--\,G0. This is consistent with the spectral type
found by S99. Our mass ratio, larger than found previously, of
$3.9\,\pm\,0.6$, along with our inclination of 68\fdg 65$\,\pm\,
1\fdg 5$ implies lower masses of $5.4\,\pm\,0.3\,$M$_{\sun}$ and
$1.45\,\pm\,0.35\,$M$_{\sun}$ for the black hole and the companion,
respectively. The mass ratio along with the $E(B-V)$ also implies a
lower luminosity of the secondary of
$21.0\,\pm\,6.0\,$L$_{\sun}$. These results are, however, rather
sensitive to the assumptions about the disc structure (even though in
$V$ the disc only contributes some 5 per cent of the light). Our
best-fitting models have a disc temperature profile that is much
flatter than that of a steady-state disc. The reduced revised masses
also help to explain the transient nature of the system

\section*{Acknowledgements}
We thank Jerry Orosz for kindly providing the light-curve
data which has made this analysis possible. MEB thanks Tariq Shahbaz
and Janet Wood for useful discussion.

\bsp

\label{lastpage}


\begin{thebibliography}{}
\bibitem{avni} Avni Y., 1976, ApJ, 210, 642
\bibitem{bailyn1} Bailyn~C.~D. et~al., 1995a, Nat, 374, 701
\bibitem{bailyn2} Bailyn~C.~D., Orosz~J.~A., McClintock~J.~E., Remillard~R.~A., 1995b, Nat, 378, 157
\bibitem{bessell} Bessell~M.~S., 1990, PASP, 102, 1181
\bibitem{bleach} Bleach~J.~N., Wood~J.~H., Catal\' an~M.~S., Welsh~W.~F., Robinson~E.~L., Skidmore~W., 2000, MNRAS, 312, 70
\bibitem{brandt} Brandt~W.~N., Podsiadlowski~Ph., Sigurdsson~S., 1995,
MNRAS, 277, L35
\bibitem{buxton} Buxton~M., Vennes~S., 2001, PASA, 18 (1), 91
\bibitem{fitzgerald} Fitzgerald~M.~P., 1970, A\&A, 4, 234
\bibitem{fitzpatrick} Fitzpatrick~E.~L., 1999, PASP, 111, 63
\bibitem{greene} Greene~J., Bailyn~C.~D., Orosz~J.~A., 2001, ApJ, 554, 1290
\bibitem{greiner} Greiner~J., Predehl~P., Pohl~M., 1995, A\&A, 297, L67
\bibitem{hameury} Hameury~J.~M., Lasota~J.~P., McClintock~J.~E., Narayan~R., 1997, ApJ, 489, 234
\bibitem{hjellming1} Hjellming~R.~M., Johnston~K.~J., 1988, ApJ,
328, 600
\bibitem{hjellming2} Hjellming~R.~M., Rupen~M.~P., 1995, Nat, 375, 464
\bibitem{horne1} Horne~K., 1993, in Wheeler~J.~C., ed., Accretion
disks in compact stellar systems. World Scientific Publishing Company,
Singapore, p. 117
\bibitem{horne2} Horne~K. et~al., 1996, IAU Circ. 6406
\bibitem{hynes} Hynes~R.~I. et~al., 1998, MNRAS, 300, 64
\bibitem{israelian} Israelian~G. et~al., 1999, Nat, 401, 142
\bibitem{johnson} Johnson~H.~L., 1966, ARA\&A, 4, 193
\bibitem{king} King~A.~R., Kolb~U., Szuskiewicz~E., 1997, ApJ, 488, 89
\bibitem{kolb1} Kolb~U., 1998, MNRAS, 297, 419
\bibitem{kolb2} Kolb~U., King~A.~R., Ritter~H., Frank~J., 1997, ApJ, 485, L33
\bibitem{kurucz} Kurucz~R.~L., 1992, Rev. Mex. Astron. Astophys. 23, 181
\bibitem{lucy} Lucy~L.~B., 1967, Z. Astrophysik, 65, 89
\bibitem{mcclintock} McClintock~J.~E., Remillard~R.~A., 1986, ApJ, 308, 110
\bibitem{mckay} McKay~D., Kesteven~M., 1994, IAU Circ. 6062
\bibitem{meyer} Meyer~F., Meyer-Hofmeister~E., 1994, A\&A, 288, 175
\bibitem{meyer-hofmeister} Meyer-Hofmeister~E., Meyer~F., 2000, A\&A, 355, 1073
\bibitem{narayan} Narayan~R., McClintock~J.~E., Yi~I., 1996, ApJ, 451, 821
\bibitem{oke} Oke~J.~B., 1977, ApJ, 217, 181
\bibitem{orosz1} Orosz~J.~A., Bailyn~C.~D., 1997, ApJ, 477, 876 (OB)
\bibitem{orosz2} Orosz~J.~A., Hauschildt~P.~H., 2000, A\&A, 364, 265
\bibitem{phillips} Phillips~S.~N., Shahbaz~T., Podsiadlowski~ Ph., 1999,
MNRAS, 304, 839 
\bibitem{podsi1} Podsiadlowski~Ph., Rappaport~S., Pfahl~E., 2001a,
ApJ, submitted
\bibitem{podsi2} Podsiadlowski~Ph., Nomoto~K., Maeda~K., Nakamura~T.,
Mazzali~P., Schmidt~B., 2001b, ApJ, submitted
\bibitem{pols} Pols~O.~R., Tout~C.~A., Schr$\ddot{\rm{o}}$der~K.,
Eggleton~P.~P., Manners~J., 1997, MNRAS, 289, 869
\bibitem{regos} Reg\H os~E., Tout~C.~A., Wickramasinghe~D., 1998,
ApJ, 509, 362
\bibitem{shahbaz1} Shahbaz~T., van~der~Hooft~F., Casares~J.,
Charles~P.~A., van~Paradijs~J., 1999, MNRAS, 306, 89 (S99)
\bibitem{shahbaz2} Shahbaz~T., Bandyopadhyay~R.~M., Charles~P.~A., 1999,
A\&A, 346, 82
\bibitem{shakura} Shakura~I.~N., Sunyaev~R.~A., 1973, A\&A, 24, 337
\bibitem{straizys} Strai$\check{\rm{z}}$ys~V., Kuriliene~G., 1981, Ap\&SS,
80, 353
\bibitem{tingay} Tingay~S.~J. et~al., 1995, Nat. 374, 101
\bibitem{tug} T\" ug~H., White~N.~M., Lockwood~G.~W., 1977,
A\&A, 61, 679
\bibitem{hooft} van~der~Hooft~F., Heemskerk~M.~H.~M., Alberts~F.,
van~Paradijs~J., 1998, A\&A, 329, 538 (vdH)
\bibitem{paradijs} van~Paradijs~J., 1996, ApJ, 464, L139
\bibitem{zeipel} von~Zeipel~H., 1924, MNRAS, 84, 665
\bibitem{wade} Wade~R.~A., Rucinski~S.~M., 1985, A\&AS, 60, 471
\bibitem{wagoner} Wagoner~R.~V., Silbergleit~A.~S.,
Ortega-Rodr\'\i guez~M., 2001, ApJL submitted (astro-ph/0107168)
\bibitem{wood1} Wood~J.~H., Horne~K., Berriman~G., Wade~R.,
O'Donoghue~D., Warner~B., 1986, MNRAS, 219, 629
\bibitem{wood2} Wood~J.~H., Horne~K., Berriman~G., Wade~R., 1989, ApJ, 341, 974
\bibitem{wood3} Wood~J.~H., Bleach~J.~N., Catal\' an~M.~S.,
Welsh~W.~F., Robinson~E.~L., 2001, in Podsiadlowski~Ph., Rappaport~S.,
King~A.~R., D'Antona~F., Burderi~L., eds, ASP Conf. Ser. Vol. 229,
Evolution of Binary and Multiple Star Systems. Astron. Soc. Pac., San
Francisco, p. 267 
\end{thebibliography}
\end{document}